\documentclass[11pt]{article}
\usepackage{acl2012}
\usepackage{times}
\usepackage{latexsym}
\usepackage{amsmath}
\usepackage{multirow}
\usepackage{url}
\usepackage{graphicx}
\usepackage[show]{chato-notes} 
\usepackage{mathptmx}
\usepackage{amsmath}
\usepackage{amssymb}
\usepackage{amsthm}
\usepackage{hyphenat}
\usepackage{subfigure}

\usepackage{cmm-greek}

\newtheorem{theorem}{Theorem}

\DeclareMathOperator*{\argmin}{arg\,min}
\newcommand{\PROBLEM}{\textsc{triangle balance}}
\newcommand{\TLSG}{\textsc{two-level spin glass}}
\newcommand{\TLSGshort}{\textsc{tlsg}}

\newcommand{\hide}[1]{}
\newcommand{\xhdr}[1]{\vspace{2mm}\noindent{{\bf #1.}}}

\newcommand{\revision}[1]{#1}

\newcommand{\tech}[1]{\emph{#1}}

\newcommand{\secref}[1]{Sec.~\ref{#1}}
\newcommand{\eqnref}[1]{Eq.~\ref{#1}}

\newcommand{\Figref}[1]{Fig.~\ref{#1}}
\newcommand{\figref}[1]{Fig.~\ref{#1}}

\newcommand{\denselist}{ \itemsep -3pt\topsep-10pt\partopsep-10pt }

\DeclareMathAlphabet{\mathcal}{OMS}{cmsy}{m}{n}

\title{Exploiting Social Network Structure for Person-to-Person\\Sentiment Analysis}

\author{Robert West \\
  Stanford University \\
  west@cs.stanford.edu
  \And
  Hristo S. Paskov \\
  Stanford University \\
  hpaskov@stanford.edu
  \And
  Jure Leskovec \\
  Stanford University \\
  jure@cs.stanford.edu
  \And
  Christopher Potts \\
  Stanford University \\
  cgpotts@stanford.edu
}

\date{}

\begin{document}
\maketitle

\begin{abstract}
Person-to-person evaluations are prevalent in all kinds of discourse
and important for establishing reputations, building social bonds, and
shaping public opinion.  Such evaluations can be analyzed separately
using signed social networks and textual sentiment analysis, but this
misses the rich interactions between language and social context.  To
capture such interactions, we develop a model that predicts individual
$A$'s opinion of individual $B$ by synthesizing information from the
signed social network in which $A$ and $B$ are embedded with sentiment
analysis of the evaluative texts relating $A$ to $B$.  We prove that
this problem is NP\hyp hard but can be relaxed to an efficiently solvable hinge-loss Markov random field, and we show that this implementation outperforms text\hyp only and network\hyp only versions in two very different datasets involving community\hyp level decision\hyp making: the Wikipedia Requests for Adminship corpus and the Convote U.S.\ Congressional speech corpus.

\end{abstract}

\section{Introduction}
\label{sec:Introduction}

People's evaluations of one another are prevalent in all kinds of
discourse, public and private, across ages, genders, cultures, and
social classes \cite{Dunbar:2004}.  Such opinions
matter
for
establishing reputations and reinforcing social bonds, and they are
especially consequential in political contexts, where they take the
form of endorsements, accusations, and assessments intended to
sway public opinion.

The significance of such \tech{person-to-person} evaluations means that
there is a pressing need for computational models and technologies
that can analyze them. Research on \tech{signed social networks}
suggests one path forward: how one person will evaluate another can
often be predicted from the network they are embedded in.
Linguistic sentiment analysis suggests another path forward: one could
leverage textual features to predict the valence of evaluative texts
describing people.  Such independent efforts have been successful, but
they generally neglect the ways in which social and linguistic
features complement each other. In some settings, textual data is
sparse but the network structure is largely observed; in others, text
is abundant but the network is partly or unreliably recorded.
In addition,
we often see rich interactions between the two kinds of
information---political allies might tease each other with negative
language to enhance social bonds, and opponents often use
sarcastically positive language in their criticisms. Separate
sentiment or signed\hyp network models will miss or misread these signals.

We develop (\secref{sec:Model}) a graphical model that synthesizes
network and linguistic information to make more and better predictions
about both.  The objective of the model is to predict $A$'s opinion of
$B$ using a synthesis of the structural context around $A$ and $B$
inside the social network and sentiment analysis of the
evaluative texts relating $A$ to $B$.  We prove that the problem is
NP\hyp hard but that it can be
relaxed to an efficiently solvable
hinge-loss Markov random field \cite{broecheler:uai10}, and we show that
this implementation outperforms text\hyp only and network\hyp only versions in
two very different datasets involving community\hyp level
decision\hyp making:
the Wikipedia Requests for Adminship corpus, in which Wikipedia editors discuss and vote on who should be promoted within the Wikipedia hierarchy (\secref{sec:Wikipedia experiments}),
and
the Convote U.S.\ Congressional speech corpus \cite{Thomas:Pang:Lee:2006}, in which elected officials discuss
political topics (\secref{sec:Convote experiments}).
These corpora differ dramatically in size, in the style and quality of their textual data, and in the structure and observability of their networks. Together, they provide a clear picture of how joint models of text and network structure can excel where their component parts cannot.

\section{Background and related work}
\label{sec:Background}


\subsection{Sentiment analysis}
\label{sec:Sentiment analysis}

In NLP, the label \tech{sentiment analysis} covers diverse phenomena
concerning how information about emotions, attitudes, perspectives,
and social identities is conveyed in language \cite{PangLee08}.  Most
work assumes a \tech{dimensional model} in which emotions are defined
primarily by valence\slash polarity and arousal\slash intensity
\cite{Russell80,Feldman:Russell:1998,Rubin:Talerico:2009}, and the
dominant application is predicting the valence of product, company,
and service reviews.

We adopt the conceptual assumptions of this work for our basic
sentiment model, but our focus is on person\hyp to\hyp person
evaluations and their social consequences.
This involves elements of work on modeling political affiliation
\cite{Agrawal-etal:2003,Malouf:Mullen:2008,Yu:Kaufmann:Diermeier:2008},
bias \cite{Yano-etal10,Recasens-etal:2013}, and stance on debate
topics
\cite{Thomas:Pang:Lee:2006,Somasundaran:Wiebe:2010,Lin-etal:2006,Anand-etal:2011},
but these aspects of belief and social identity are not our primary
concern.  Rather, we expect them to be \emph{predictive} of the
sentiment classifications we aim to make---e.g., if two people share
political views, they will tend to evaluate each other positively.

Recent work in sentiment analysis has brought in topical, contextual,
and social information to make more nuanced predictions about language
\cite{Jurafsky-etal:2014,Wilson-etal05,Blitzer:Dredze:Pereira:2007}.
We build on these insights with our model, which seeks to modulate
sentiment predictions based on network information (and \textit{vice
  versa}).


\subsection{Signed-network analysis}
\label{sec:Social network analysis}

\begin{figure*}
 \centering
\begin{tabular}{c|c}
 	\subfigure[Theories of social balance and status]{
	    \hspace{-5mm}
		    \includegraphics[width=.66\textwidth]{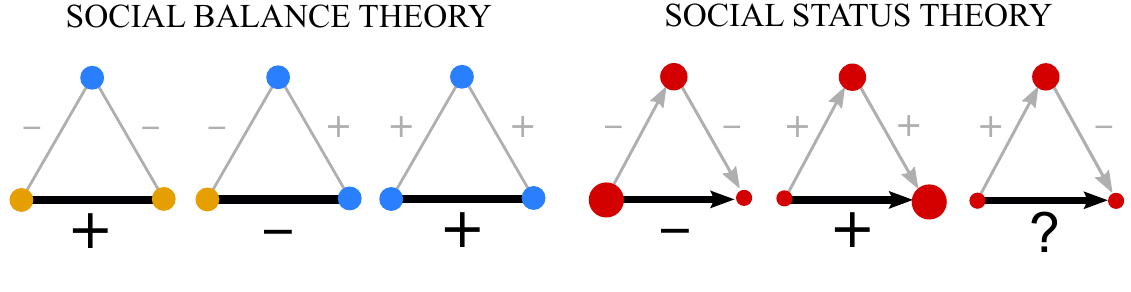}
            \label{fig:status_balance}
	}
	&
 	\subfigure[Desiderata]{
		    \includegraphics[width=.32\linewidth]{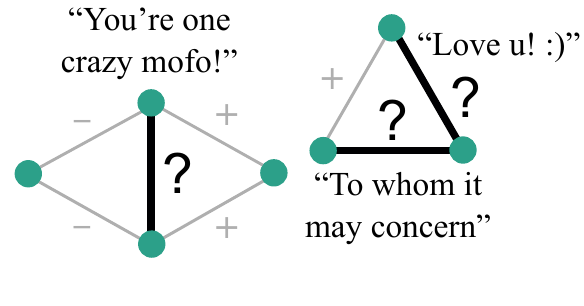}
            \label{fig:desiderata}
	}
	\end{tabular}
	\vspace{-5mm}
            \caption{
            \textbf{(a)}
            		Predictions of social balance and status theories for the bold black edge, given the thin gray edges.
				Balance theory reasons about undirected, status theory about directed, triangles.
				In the status diagrams, node size signifies social status.
				A positive edge may be replaced by a negative edge in the opposite direction, and \textit{vice versa}, without changing the prediction.
				Status theory makes no prediction in the rightmost case.
            \textbf{(b)} Situations of the sort we aim to capture. At left, the network
            resolves textual ambiguity. At right, the text compensates for edge\hyp label sparsity.
			}
 \label{fig:status_balance_desirata}
\end{figure*}

Many social networks encode person-to-person sentiment information via
\tech{signed} edges between users summarizing their opinions of each
other. In this setting, one can leverage sociological theories of
pairwise relationships and group-level organization to identify and
understand patterns in these relationships \cite{Heider:1946,Cartwright:Harary:1956}.

\tech{Balance theory} is based on simple intuitions like `a friend of
my friend is my friend', `an enemy of my enemy is my friend', and
`an enemy of my friend is my enemy'.  In graph theory, these are
statements about the edge signs of triangles of connected nodes: given
the signs of two edges, balance theory predicts the third, as
summarized in Fig.~\ref{fig:status_balance}, where the two given edges
(gray) determine the third
(black). 

For directed relationships, \newcite{leskovec2010signed} formulate an
alternative called \tech{status theory,} which posits that networks
organize according to social status: a node has positive edges to
others with higher status and negative edges to those with lower
status. Fig.~\ref{fig:status_balance} illustrates the structure of
various directed signed triangles, where the sign of the third 
edge (black)  can be inferred based on the signs and directions of the other
two (gray).

\newcite{leskovec2010signed} show that signed edges in networks emerge
in a manner that is broadly consistent with both of these theories
and that social\hyp network
structure alone can support accurate edge-sign predictions
\cite{leskovec2010predicting}.
\newcite{kunegis2013added} predict hidden positive and negative edges in a scenario where
all observed edges are positive.
\newcite{bach2013hinge} and \newcite{huang2013flexible}
frame sign prediction as a hinge-loss
Markov random field, a type of probabilistic graphical model
introduced by \newcite{broecheler:uai10}.
Our model combines these ideas
with a sentiment model to achieve even more robust predictions.

\subsection{Synthesis of sentiment \& network analysis}
\label{sec:social-sentiment}

Models of sentiment and signed networks have been successful at a
variety of tasks.  However, missing from the current scientific
picture is a deep understanding of the ways in which sentiment
expression and social networks interact. 
To some extent, these interactions are captured by adding contextual
and demographic features to a text-based sentiment model, but those
features only approximate the rich relational structure encoded in a
signed network.

\newcite{Thomas:Pang:Lee:2006} and \newcite{Tan-etal:2011} capitalize
on this insight using an elaboration of the graph-cuts approach of
\newcite{PangLee04}. They are guided by an assumption of
\tech{homophily,} i.e., that certain social relationships correlate
with agreement on certain topics: \newcite{Thomas:Pang:Lee:2006} use
party affiliation and mentions in speeches to predict voting patterns,
and \newcite{Tan-etal:2011} use Twitter follows and mentions to
predict attitudes about political and social events.  Related ideas are
pursued by \newcite{ma2011recommender} and \newcite{hu2013exploiting},
who add terms to their models enforcing homophily between friends with
regard to their preferences.

We adopt some of the assumptions of the above authors, but our task is
fundamentally different in two respects.  First, whereas they model
person-to-item evaluations, we model person-to-person evaluations;
these are also the focus of \newcite{Tang:2013:EHE:2433396.2433405},
who, though, use an unsigned network, 
whereas our work is geared toward distinguishing positive and negative
edge labels. Second, the above models make overarching homophily
assumptions, whereas we allow our model to explore the full set of
triangle configurations suggested by Fig.~\ref{fig:status_balance}.

\section{Model}
\label{sec:Model}

Here, we argue that combining textual and structural features can help predict edge signs.
We formulate a model, show that it is computationally hard, and provide a relaxed version that is computationally tractable, building on work by \newcite{bach2013hinge}.

\subsection{Desiderata}
\label{sec:Desiderata}

In many real-world scenarios, rich features are associated with edges between two people, such as comments they made about each other, messages they exchanged, or other behavioral features.
Such features may contain a strong sentiment signal useful for predicting edge signs and may be used to fit a conventional sentiment model (\secref{sec:Sentiment analysis}).

However, the sign of an edge also depends on the signs of surrounding edges
in the network
(\secref{sec:Social network analysis}).
A purely edge\hyp feature--based sentiment model cannot account for the network structure, since it reasons about edges as independent of each other.


We argue that considering sentiment and network structure jointly can result in better predictions than either one on its own.
\Figref{fig:desiderata} provides two illustrative examples.
Here, the gray edge signs are observed, while the polarities of the black edges are to be predicted.
In the left network, the text of the black edge (\textit{`You're one crazy mofo!'}) might suggest a negative polarity.
However, a negative black edge would make both triangles
inconsistent with balance theory
(\Figref{fig:status_balance}), whereas a positive black edge makes
them consistent with the theory.
So, in this case, the network context effectively helps detect the teasing, non\hyp literal tone of the statement.

In the right network of Fig.~\ref{fig:desiderata}, only one of three edge signs is observed.
Predicting two positive edges would be consistent with balance theory, but the same would be true for predicting two negative edges.
The text on the lower black edge
\revision{(\textit{`To whom it may concern'})}
does not carry any
clear
sentiment signal, but the \textit{`Love u!\ :)'} on the other edge strongly suggests a positive polarity.
This
lets us conclude that the bottom edge should probably be positive, too, since otherwise the triangle would contradict balance theory.
This
shows that combining sentiment and network features can
help
when jointly reasoning about several unknown edge signs.

\subsection{Problem formulation}
\label{sec:Problem formulation}

We now formulate a model capable of synthesizing textual and network features.

\xhdr{Notation}
We represent the given social network as a signed graph $G = (V, E, x)$, where the vertices $V$ represent people; the edges $E$, relationships between people in $V$; and the sign vector $x \in \{0,1\}^{|E|}$ represents edge polarities, i.e., $x_e=1$ ($x_e=0$) indicates a positive (negative) polarity for edge $e \in E$.

Some types of relationships imply directed edges (e.g., following a user on Twitter, or voting on a candidate in an election), whereas others imply undirected edges (e.g., friendship on Facebook, or agreement in a network of politicians).
We formulate our problem for undirected graphs here,
but the extension to directed graphs is straightforward.
We define a \emph{triangle} $t=\{e_1,e_2,e_3\}\subseteq E$ to be a set of three edges that form a cycle, and use $T$ to indicate the set of all triangles in $G$.
Finally, we use $x_t = (x_{e_1},x_{e_2},x_{e_3}) \in \{0,1\}^3$ to refer to $t$'s edge signs.

\xhdr{Optimization problem}
We assume that the structure of the network (i.e., $V$ and $E$) is fully observed, whereas the edge signs $x$ are only partially observed.
Further, we assume that we have a sentiment model that outputs, for each edge $e$ independently, an estimate $p_e$ of the probability that $e$ is of positive polarity, based on textual features associated with $e$.
The task, then, is to {\em infer the unobserved edge signs} based on the observed information.

The high\hyp level idea is that we want to infer edge signs that (1)~agree with the predictions of the sentiment model, and (2)~form triangles that agree with social theories of balance and status.
It is not always possible to meet both objectives simultaneously for all edges and triangles, so we need to find a trade-off.
This gives rise to a combinatorial optimization problem, which we term \PROBLEM,
that seeks to find edge signs $x^*$ that minimize an objective consisting of both edge and triangle costs:%
\footnote{Of course, the entries of $x^*$ corresponding to observed edges are not variable but fixed to their observed values in \eqnref{eqn:objective}.}
\begin{eqnarray}
x^* \;\; = \;\; \argmin_{x \in \{0,1\}^{|E|}} \;\;\; \sum_{e \in E} c(x_e, p_e) + \sum_{t \in T} d(x_t).
\label{eqn:objective}
\end{eqnarray}
The first term is the \emph{total edge cost,} in which each edge $e$ contributes a cost capturing how much its inferred sign $x_e$ deviates from the prediction $p_e$ of the sentiment model.
The second term, the \emph{total triangle cost,} penalizes each triangle $t$ according to how undesirable its configuration is under its inferred signs $x_t$ (e.g., if it contradicts status or balance theory).

We use the following edge cost function:
\begin{eqnarray}
c(x_e, p_e) \;\; = \;\; \lambda_1(1-p_e)x_e + \lambda_0 p_e(1-x_e).
\label{eqn:c}
\end{eqnarray}
Here, $\lambda_1, \lambda_0\in\mathbb{R_+}$ are tunable parameters that allow for asymmetric costs for positive and negative edges, respectively, and $p_e$ is the probability of edge $e$ being positive according to the sentiment model alone. Intuitively, the more the inferred edge sign $x_e$ deviates from the prediction $p_e$ of the sentiment model, the higher the edge cost. 
(Note that at most one of the two sum factors of \eqnref{eqn:c} is non\hyp zero.)

The triangle cost for triangle $t$ is signified by $d(x_t)$, which can only take on 8 distinct values because $x_t \in \{0,1\}^3$
(in practice, there are symmetries that decrease this number to 4).
The parameters $d(x_t)$ may be tuned so that triangle configurations that agree with social theory have low costs, while those that disagree with it (e.g., `the enemy of my friend is my friend') have high costs.

\subsection{Computational complexity}
\label{sec:Computational complexity}

The problem defined in \eqnref{eqn:objective} is intuitive, but, as with many combinatorial optimization problems, it is hard to find a good solution.
In particular, we sketch a proof of this theorem in Appendix~\ref{sec:proof}:

\begin{theorem}
\label{thm:np-completeness}
\PROBLEM{} is NP\hyp hard.
\end{theorem}

\subsection{Relaxation as a Markov random field}
\label{sec:Convex relaxation}

\revision{
The objective function of \eqnref{eqn:objective} may be seen as defining a Markov random field (MRF) over the underlying
social
network $G$, with edge potentials (defined by $c$) and triangle potentials (defined by $d$).
Inference in MRFs (i.e., computing $x^*$) is a well\hyp studied task for which a variety of methods have been proposed \cite{koller2009probabilistic}.
However, since our problem is NP-hard, no method can be expected to find $x^*$ efficiently.
One way of dealing with the computational hardness would be to find an \emph{approximate binary} solution, using techniques such as Gibbs sampling or belief propagation.
Another option is to consider a continuous relaxation of the binary problem and find an \emph{exact
non\hyp binary} solution whose edge signs are continuous, i.e., $x_e\in[0,1]$.

We take this latter approach and cast our problem as a \textit{hinge\hyp loss Markov random field} (HL-MRF). This is inspired by \newcite{bach2013hinge}, who also use an HL-MRF to predict edge signs based on triangle structure, but do not use any edge features.
An HL-MRF is an MRF with continuous variables and with potentials that can be expressed as sums of hinge-loss terms of linear functions of the variables (cf.\ \newcite{broecheler:uai10} for details).
HL-MRFs have the advantage that their objective function is convex so that, unlike binary MRFs (as defined by \eqnref{eqn:objective}), exact inference is efficient \cite{bach2013hinge}.

We achieve a relaxation by using sums of hinge\hyp loss terms to interpolate $c$ over the continuous domain $[0,1]$ and $d$, over $[0,1]^3$ (even though they are defined only for binary domains).
As a result, the HL-MRF formulation is equivalent to \eqnref{eqn:objective} when all $x_e$ are binary, but it also handles continuous values gracefully.
We now interpret a real-valued `sign' $x_e \in [0,1]$ as the degree to which $e$ is positive.

We start by showing how to transform $c$:
even though it could be used in its current form (\eqnref{eqn:c}), we create a tighter relaxation by using
\begin{eqnarray}
\tilde{c}(x_e,p_e) = \lambda_1\|x_e - p_e\|_+ +  \lambda_0\|p_e - x_e\|_+,
\label{eqn:c_tilde}
\end{eqnarray}
where $\|y\|_+=\max\{0,y\}$ is the hinge loss.
At most one term can be active for any $x_e\in[0,1]$ due to the hinge loss, and $c(x_e,p_e)=\tilde{c}(x_e,p_e)$ for binary $x_e$. 

To rewrite $d$, notice that, for any $x_t\in\{0,1\}^3$, we can write $d$ as
\begin{eqnarray}
d(x_t) = \sum_{z\in \{0,1\}^3} d(z)\; \delta(x_t,z),
\label{eqn:d_sum}
\end{eqnarray}
where $\delta(x_t,z) = 1$ if $x_t=z$ and $0$ otherwise. While $\delta$ is not convex, we can use
\begin{eqnarray}
f(x_t,z) = \left\|1 - \|x_t - z\|_1\right\|_+
\label{eqn:f}
\end{eqnarray}
as a convex surrogate. When $x_t$ is binary, either $x_t=z$ so $\|x_t - z\|_1=0$ or $x_t\neq z$ so $\|x_t - z\|_1\geq 1$, and hence $f(x_t,z) = \delta(x_t,z)$.
To prove convexity, note that, for any fixed binary $z \in \{0,1\}^3$,
$\|x-z\|_1 = \sum_{i=1}^3 |x_i-z_i|$
is linear in $x \in [0,1]^3$, since $|x_i-z_i|$ equals either $x_i$ (if $z_i=0$) or $1-x_i$ (if $z_i=1$).
It follows that $f$ is a hinge\hyp loss function of a linear transformation of $x_t$ and therefore convex in $x_t$.

Requiring the triangle cost $d(z)$ to be nonnegative for all triangle types $z\in\{0,1\}^3$, we can use
\begin{eqnarray}
\tilde{d}(x_t) = \sum_{z\in \{0,1\}^3}d(z)\; f(x_t,z)
\end{eqnarray}
as a convex surrogate for $d$.
Our overall optimization problem is then the following relaxation of \eqnref{eqn:objective}:
\begin{eqnarray}
x^* \;\; = \;\; \argmin_{x \in [0,1]^{|E|}} \;\;\; \sum_{e \in E} \tilde{c}(x_e, p_e) + \sum_{t \in T} \tilde{d}(x_t).
\label{eqn:objective_relaxed}
\end{eqnarray}
This objective has the exact form of an HL-MRF, since it is a weighted sum of hinge losses of linear functions of $x$.
We use the \textit{Probabilistic Soft Logic} package%
\footnote{http://psl.umiacs.umd.edu}
to perform the optimization, which is in turn based on the alternating\hyp direction method of multipliers (ADMM) \cite{boyd2011distributed}.
}



\xhdr{Learning}
Clearly, a solution is only useful if the cost parameters ($\lambda_1$, $\lambda_0$, and $d(z)$ for all $z \in \{0,1\}^3$) are set appropriately.
One option would be to set the values heuristically, based on the predictions made by the social balance and status theories (\secref{sec:Social network analysis}).
\revision{However, it is more principled to learn these parameters from data.
For this purpose, we leverage the learning procedures included in the HL-MRF implementation we employ, which uses the voted\hyp perceptron algorithm to perform maximum\hyp likelihood estimation \cite{bach2013hinge}.
}

\begin{figure}
 \centering
    \includegraphics[width=\linewidth]{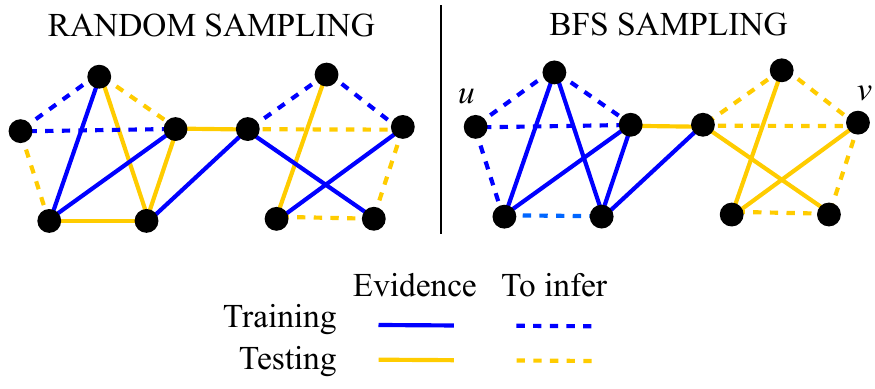}
    \vspace{-8mm}
 \caption{
	Options for training and testing our model.
	}
 \label{fig:train_test}
\end{figure}

Since our data points (edges) interact with each other via the network, some words on how we perform training and testing are in order.
Fig.~\ref{fig:train_test} shows two options for obtaining training and testing sets (we use both options in our experiments).
In the `random sampling' paradigm, we randomly choose a set of edges for training (blue), and a disjoint set of edges for testing (yellow).
In `BFS sampling', we run a breadth-first search from seed node $u$ to obtain a coherent training set (blue), and likewise from a seed node $v$ to obtain a coherent testing set (yellow), taking care that no edges from the training set are also included in the testing set.

During both training and testing, an arbitrary portion of the edge signs may be fixed to observed values and need not be inferred.
These are the solid edges in Fig.~\ref{fig:train_test}; we refer to them as \textit{evidence.}
Further, we define the \textit{evidence ratio} as the number of evidence edges, divided by the number of all edges considered (solid and dashed).

The learning algorithm may use the structure ($V$ and $E$) of the training graph induced by all blue edges (solid and dashed), the predictions $p_e$ of the sentiment model for all blue edges, and the signs of the solid blue edges to predict the dashed blue edges.

During testing, the network structure of all yellow edges, the sentiment predictions for all yellow edges, and the signs of the solid yellow edges may be used to predict the dashed yellow edge signs.
In principle, all training edges could be used as extra evidence for testing (i.e., all blue edges may be made solid yellow).
However, in our experiments, we keep the training and testing sets fully disjoint.

\xhdr{Technical details}
For clarity, we give further details.
First, the distribution of positive and negative signs may be skewed;
e.g., we observe a prior probability of 76\% positive signs in our Wikipedia corpus (\secref{sec:Wikipedia experiments}).
Therefore,
as also done by \newcite{bach2013hinge},
we add a cost term to our objective (\eqnref{eqn:objective_relaxed}) that penalizes deviations from this prior probability (as estimated on the training set).
This ensures that the model can default to a reasonable prediction for edges
that are
not embedded in any triangles
and about which the sentiment model is uncertain.

Second, intuitively, we should not penalize deviating from the sentiment model when it is itself uncertain about its prediction (i.e., when $p_e$ is far from both 0 and 1).
Rather, we want to rely more heavily on signals from the network structure in such cases.
To achieve this, we introduce 10 pairs of cost parameters $(\lambda_1^{(1)},\lambda_0^{(1)}), \dots, (\lambda_1^{(10)},\lambda_0^{(10)})$.
Then, we divide the interval $[0,1]$ into 10 bins, and when $p_e$ falls into the $i$-th bin, we use $\lambda_1^{(i)}$ and $\lambda_0^{(i)}$ in \eqnref{eqn:c_tilde}.
This way, larger costs can be learned for the extreme bins close to 0 and 1 than for the intermediate bins around 0.5.

\revision{
Finally, hinge-loss terms may optionally be squared in HL-MRFs.
We use the squared hinge loss in \eqnref{eqn:f}, since initial experimentation showed this to perform slightly better  than the linear hinge loss.
}

\section{Wikipedia experiments}
\label{sec:Wikipedia experiments}

Our first set of experiments is conducted on the Wikipedia Requests for Adminship corpus, which allows us to evaluate our model's ability to predict person\hyp to\hyp person evaluations in Web texts that are informal but
pertain to important social outcomes.

\subsection{Dataset description}
\label{sec:Wikipedia dataset}

For a Wikipedia editor to become an administrator, a \textit{request for adminship} (RfA)%
\footnote{http://en.wikipedia.org/wiki/Wikipedia:RfA}
must be submitted, either by the candidate or by another community member.
Subsequently, any Wikipedia member may cast a supporting, neutral, or opposing vote.
This induces a directed, signed network in which nodes represent Wikipedia members and edges represent votes (we discard neutral votes).%

We crawled and parsed all votes since the adoption of the RfA process in 2003 through May 2013.%
\footnote{Data available online \cite{project-website}.}
This signed network was previously analyzed by Leskovec et al.\ \shortcite{leskovec2010signed,leskovec2010predicting}.
However, there is also a rich textual component that has so far remained untapped for edge-sign prediction:
each vote is typically accompanied by a short comment (median\slash mean: 19\slash 34 tokens).
A typical positive comment reads, \textit{`I've no concerns, will make an excellent addition to the admin corps',}
while an example of a negative comment is, \textit{`Little evidence of collaboration with other editors and limited content creation.'}
The presence of a voting network alongside textual edge features makes our method of \secref{sec:Model} well\hyp suited for this dataset.

The RfA network contains 11K nodes, 160K edges (76\% positive), and close to 1M triangles.


\subsection{Experimental setup}
\label{sec:Wikipedia experimental setup}

\xhdr{Train/test sets}
We follow the train--test paradigm termed `BFS sampling' in \secref{sec:Convex relaxation} and \figref{fig:train_test}, choosing 10 random seed nodes, from each of which we perform a breadth-first search (following both in- and out-links) until we have visited 350 nodes.
We thus obtain 10 subgraphs with 350 nodes each.
We train a model for each subgraph $i$ and test it on subgraph $i+1$ (mod 10), ensuring that edges from the training graph are removed from the testing graph.

\revision{
The BFS sampling paradigm was used because the alternative (`random sampling' in \figref{fig:train_test}) produces subgraphs with mostly
isolated edges and only a few triangles---an unrealistic scenario.
}

\xhdr{Evaluated models}
We evaluate three models:
\begin{enumerate}
\denselist
\item A standard, text-based sentiment model that treats edges as independent data points;
\item our full model as specified in \secref{sec:Convex relaxation}, which combines edge costs based on the predictions of the text-based sentiment model with triangle costs capturing network context;
\item a version of our model that considers only triangle costs, while ignoring the predictions of the text-based sentiment model (akin to the model proposed by \newcite{bach2013hinge}).
\end{enumerate}
We refer to these models as \textit{`sentiment',} \textit{`combined',} and \textit{`network',} respectively.

\xhdr{Sentiment model}
Our text-based sentiment model is an $L_2$\hyp regularized logistic\hyp regression classifier whose features are term frequencies of the 10,000 overall most frequent words.
The $L_2$-penalty is chosen via cross-validation on the training set.
Since comments often explicitly contain the label (\textit{`support'} or \textit{`oppose'}), we remove all words with prefixes \textit{`support'} or \textit{`oppos'}.
We train the model only once, on a random sample of 1,000 comments drawn from the set of all
160K
comments (the vast majority of which will not appear in our 10 subgraphs).

\xhdr{Evidence ratio}
Regarding the other two models, recall from \secref{sec:Convex relaxation} our definition of the \textit{evidence ratio,} the fraction of edge signs that are fixed as evidence and need not be inferred.
In our experiments, we explore the impact of the evidence ratio during training and testing, since we expect performance to increase as more evidence is available. (We use the same evidence ratio for training and testing, but this need not necessarily be so.)

\xhdr{Metrics}
As our principal evaluation metrics, we use the areas under the curve (AUC) of the receiver operating characteristic (ROC) curve as well as the precision--recall (PR) curves.
There are two PR curves, one for the positive class, the other for the negative one.
Of these two, the positive class is less interesting: due to the class imbalance of 76\% positive edges, even a random guesser would achieve an AUC of 0.76.
The PR curve of the negative class is more informative: here, it is much harder to achieve high AUC, since random guessing yields only 0.24.
Moreover, the negative edges are arguably more important, not only because they are rarer, but also because they indicate tensions in the network, which we might be interested in detecting and resolving.
For these reasons, we report only the AUC under the negative PR curve (AUC\slash negPR) here.

Additionally, we report the area under the ROC curve (AUC\slash ROC), a standard metric for quantifying classification performance on unbalanced data. It captures the probability of a random positive test example receiving a higher score than a random negative one (so guessing gives an AUC\slash ROC of 0.5).


\subsection{Results}
\label{sec:Wikipedia results}

\begin{figure}
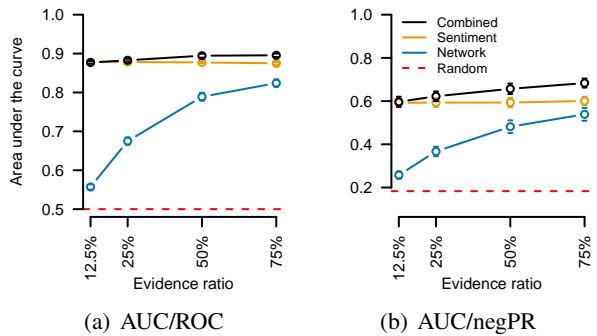

 \centering
 	\subfigure[AUC/ROC]{
	    \hspace{-2mm}
	    \includegraphics[scale=1]{{{FIG/wiki_rfa/auc-roc_varyingObsRatio_discard=0_interpolated_minEmbed=0}}}
		\label{fig:auc-roc_wiki_varyingObsRatio}
	}
 	\subfigure[AUC/negPR]{
	    \includegraphics[scale=1]{{{FIG/wiki_rfa/auc-pr_neg_varyingObsRatio_discard=0_interpolated_minEmbed=0}}}
		\label{fig:auc-pr_neg_wiki_varyingObsRatio}
	}
     \vspace{-5mm}
\caption{
	AUC as function of evidence ratio (Wikipedia), with standard errors.
	}
 \label{fig:auc_wiki_varyingObsRatio}
     \vspace{-2mm}
\end{figure}

\xhdr{Performance as a function of evidence ratio}
The AUCs as functions of the evidence ratio are shown in \figref{fig:auc-roc_wiki_varyingObsRatio}.
(We emphasize that these plots are not themselves ROC and PR curves; rather, they are derived from those curves by measuring AUC for a range of models, parametrized by evidence ratio.)

Since we use the same sentiment model in all cases (\secref{sec:Wikipedia experimental setup}), its performance (yellow) does not depend on the evidence ratio.
It is remarkably high, at an AUC/ROC of 0.88, as a consequence of the highly indicative, sometimes even formulaic, language used in the comments (examples in \secref{sec:Wikipedia experimental setup}).

The network\hyp only model (blue) works poorly on very little evidence (AUC/ROC 0.56 for 12.5\% evidence) but improves steadily as more evidence is used (AUC/ROC 0.82 for 75\% evidence): this is intuitive, since more evidence means stronger context for each edge sign to be predicted.

Although the network\hyp only model works poorly on little evidence, our full model (black), which synthesizes the sentiment and network models, is not affected by this and effectively defaults to the behavior of the sentiment\hyp only model.
Furthermore, although the network\hyp only model never attains the performance of the sentiment\hyp only model, combining the two in our full model (black) nonetheless yields a small performance boost in terms of AUC/ROC to 0.89 for 75\% evidence.
The gains are significantly larger when we consider AUC/negPR instead of AUC/ROC:
while the sentiment model achieves 0.60, the combined model improves on this by 13\%, to 0.68, at 75\% evidence ratio.

\begin{figure}
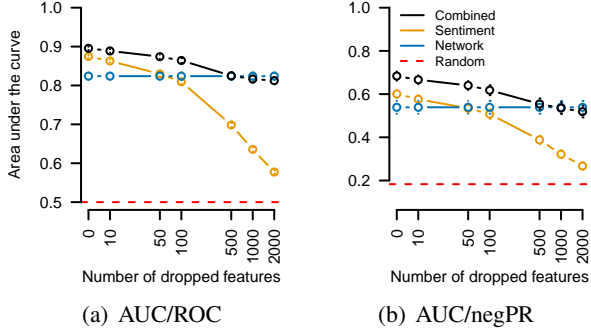

 \centering
 	\subfigure[AUC/ROC]{
	    \hspace{-2mm}
	    \includegraphics[scale=1]{{{FIG/wiki_rfa/auc-roc_varyingDiscardedFeats_obsRatio=0.75_interpolated_minEmbed=0}}}
		\label{fig:auc-roc_wiki_varyingDiscardedFeats}
	}
 	\subfigure[AUC/negPR]{
	    \includegraphics[scale=1]{{{FIG/wiki_rfa/auc-pr_neg_varyingDiscardedFeats_obsRatio=0.75_interpolated_minEmbed=0}}}
		\label{fig:auc-pr_neg_wiki_varyingDiscardedFeats}
	}
    \vspace{-5mm}
\caption{
    \hspace{-2mm}
	AUC as function of number of dropped features (Wikipedia), with standard errors.
	Evidence ratio 75\%.
	}
 \label{fig:auc_wiki_varyingDiscardedFeats}
\end{figure}

\xhdr{Performance as a function of sentiment\hyp model quality}
It seems hard to improve by much on a sentiment model that achieves an AUC/ROC of 0.88 on its own;
the Wikipedia corpus offers an exceptionally explicit linguistic signal.
Hence, in our next experiment, we explore systematically how our model behaves under a less powerful sentiment model.

First, we measure, for each feature (i.e., word), how informative it is on its own for predicting the signs of edges (quantified by its mutual information with the edge sign), which induces a ranking of features in terms of informativeness.
Now, to make the sentiment model less powerful in a controlled way, we drop the top $m$ features and repeat the experiment described above for a range of $m$ values (where we keep the evidence ratio fixed at 75\%).


\Figref{fig:auc_wiki_varyingDiscardedFeats} shows that the performance of the sentiment model (yellow) declines drastically as more features are removed.
The combined model (black), on the contrary, is much less affected: when the performance of the sentiment model drops to that of the network\hyp only (blue; ROC/AUC 0.81), the combined model is still stronger than both (0.86).
Even as the sentiment model approaches random performance, the combined model still never drops below the network\hyp only model---it simply learns to disregard the predictions of the sentiment model altogether.

\begin{figure}
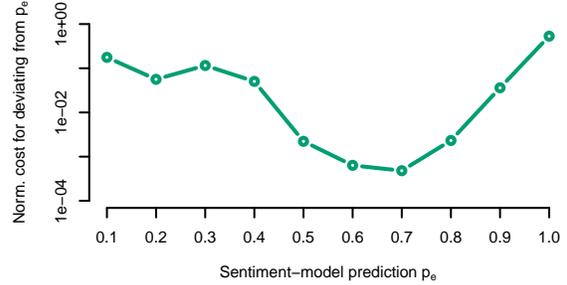

 \centering
    \includegraphics[scale=1]{{{FIG/wiki_rfa/weights_logreg-bins_varyingObsRatio_discard=0_obsRatio=0.75}}}
     \vspace{-3mm}
 \caption{
	Normalized cost $\lambda^{(i)}$ (defined in \secref{sec:Wikipedia results}; logarithmic scale) for deviating from sentiment\hyp model predictions $p_e$, for bins $i=1,\dots,10$ (Wikipedia). Upper bin boundaries on the $x$-axis.
	Values shown are averages over 10 folds.
	Evidence ratio 75\%.
	}
 \label{fig:logreg_costs}
\end{figure}

\xhdr{Learned edge costs}
Recall from the final part of Sec.\ \ref{sec:Convex relaxation} that each output $p_e$ of the sentiment model falls into one of 10 bins $[0,0.1], \dots, [0.9,1]$, with separate edge\hyp cost parameters
$\lambda_1^{(i)}$ and $\lambda_0^{(i)}$
learned for each bucket $i$.
The rationale was to give the model the freedom to trade off edge and triangle costs differently for each edge $e$, depending on how informative the sentiment model's prediction $p_e$ is.

The goal of this section is to understand whether our model indeed exposes such behavior.
Recall from Eq.~\ref{eqn:c_tilde}
that $\lambda_1$ is the constant with which the absolute difference between $p_e$ and the inferred edge sign $x_e$ is multiplied when $x_e>p_e$, while $\lambda_0$ is the constant when $x_e<p_e$.
If $p_e$ falls into bin $i$, the sum $\lambda_1^{(i)} + \lambda_0^{(i)}$ expresses the cost of deviating from $p_e$ in a single number;
further, dividing this sum by the sum of all costs (i.e., $\lambda_1^{(i)}$ and $\lambda_0^{(i)}$ for all bins $i$, plus the costs $d(z)$ of all triangle types $z$) yields a normalized edge cost for each bin $i$, which we call $\lambda^{(i)}$.

Fig.~\ref{fig:logreg_costs} plots $\lambda^{(i)}$ for all bins $i=1,\dots,10$.
We observe that deviating from the sentiment model costs more when it makes a strong prediction (i.e., $p_e$ close to 0 or 1) than when it makes a non\hyp informative one (e.g., $p_e \approx 0.5$).
When $p_e \approx 1$, nearly 100\% of the total cost is spent to ensure $x_e \approx p_e$, whereas that fraction is only around 0.1\% when $p_e \approx 0.6$.

\xhdr{Leave-one-out setting}
Our model predicts many missing edge signs simultaneously, using joint inference.
Another scenario was proposed by \newcite{leskovec2010predicting}, who predict signs one at a time, assuming all other edge signs are known.
We call this a \textit{leave-one-out} (`LOO' for short) setting.
Assume we want to predict the sign of the edge $(u,v)$ in the LOO setting, and that $u$, $v$, and $w$ form a triangle. The type of this triangle can be described by the directions and known polarities of the two edges linking $u$ to $w$, and $v$ to $w$, respectively.
The edge $(u,v)$ may be embedded in several triangles, and the histogram over their types then serves as the feature vector of $(u,v)$ in a logistic regression;
as additional features, counts of $u$'s positive\slash negative out-links and $v$'s positive\slash negative in-links are used.

\begin{figure}
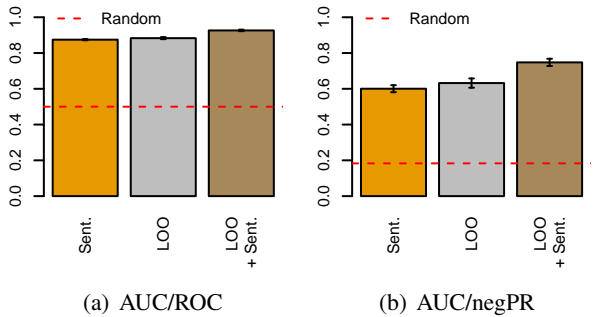

 \centering
 	\subfigure[AUC/ROC]{
	    \hspace{-2mm}
	    \includegraphics[scale=1]{{{FIG/wiki_rfa/auc_leave-one-out_roc_obsRatio=0.75}}}
		\label{fig:auc-roc_wiki_leave-one-out}
	}
 	\subfigure[AUC/negPR]{
	    \includegraphics[scale=1]{{{FIG/wiki_rfa/auc_leave-one-out_pr_neg_obsRatio=0.75}}}
		\label{fig:auc-pr_neg_wiki_leave-one-out}
	}
     \vspace{-5mm}
\caption{
	Performance in a leave\hyp one\hyp out (LOO) scenario (Wikipedia), with standard errors.
	For comparison: performance of the sentiment model alone.
	}
 \label{fig:auc_wiki_leave-one-out}
     \vspace{-3mm}
\end{figure}

Since predictions in the LOO setup can draw on the full triangle neighborhood of the edge in question, we expect it to perform better than the network\hyp only model in which edge signs in the triangle neighborhood are often missing.
This expectation is confirmed by \figref{fig:auc_wiki_leave-one-out}, which shows that the LOO model (gray) achieves an AUC/ROC (AUC/negPR) of 0.88 (0.63), with the network\hyp only model (\figref{fig:auc_wiki_varyingObsRatio}) at just 0.82 (0.54) at 75\% evidence ratio.

However, LOO is outperformed by our combined model incorporating sentiment information (\figref{fig:auc_wiki_varyingObsRatio}),
which attains an AUC/ROC (AUC/negPR) of 0.89 (0.68).
Finally, when we add the sentiment prediction as another feature to the LOO model (`LOO + Sent.' in Fig.~\ref{fig:auc_wiki_leave-one-out}), we do best, at 0.93 (0.75).

To summarize, we make two points:
(1)~By combining sentiment and network features, our model achieves better performance than a network\hyp only model (LOO) that has access to significantly more network information.
(2)~Incorporating sentiment information helps not only in our setup as described in \secref{sec:Wikipedia experimental setup}, but also in the previously proposed leave\hyp one\hyp out setup \cite{leskovec2010predicting}.


\section{U.S.\ Congress experiments}
\label{sec:Convote experiments}

We now evaluate our model in a setting in which the linguistic person\hyp to\hyp person evaluations are less direct and reliable than in the RfA corpus but the signed network is considerably denser.

\subsection{Dataset description}
\label{sec:Convote dataset}

The `Convote' corpus of Congressional speeches \cite{Thomas:Pang:Lee:2006}
consists of 3,857 speech segments drawn
from 53 debates from the U.S.\ House of Representatives in 2005.
There is a mean of 72.8 speech segments per debate and 32.1
speakers per debate. Segments are annotated with the speaker,
their party affiliation, the bill discussed, and how the
speaker voted on that bill (positive or negative).

\newcite{Thomas:Pang:Lee:2006} and others represent this corpus as a
bipartite person--item graph with signed edges from Congresspeople to the
bills (items) they spoke about, and they add additional person--person
edges encoding who mentioned whom in the speech segments.  We take a
different perspective, extracting from it a dense, undirected
person--person graph by linking two Congresspeople if they ever voted on
the same bill, labeling the edge as positive if they cast the same
vote at least half of the time.
We directly use the sentiment model trained by Thomas et al.
The resulting graph has 276 nodes,
14,690 edges (54\% positive), and 506,327 triangles.

\subsection{Experimental setup}
\label{sec:Convote experimental setup}

We split the network $G=(V,E)$ into 5 folds using the `random
sampling' technique described in \secref{sec:Convex relaxation} and
\figref{fig:train_test}: the set of nodes $V$ is fixed
across all folds, and the set of edges $E$ is partitioned randomly so
that each fold has 20\% of all edges. In the full graph, there is one
clique per debate, so each fold contains
the overlay of several subgraphs, one per debate and each 20\% complete on average.

\revision{
Here, random sampling was used because the alternative (`BFS sampling'
in \figref{fig:train_test}) would produce nearly complete subgraphs,
on which we found the prediction task to be overly easy (since the problem
becomes more constrained; \secref{sec:Convote results}).
}

We compare the three models also used on the Wikipedia dataset
(\secref{sec:Wikipedia experimental setup}).
Our sentiment model comes right out of the box with the Convote
corpus: \newcite{Thomas:Pang:Lee:2006} distribute the text-level
scores from their SVM classifier with the corpus,
so we simply work with those, after transforming them into probabilities
via logistic regression (a standard technique called Platt scaling \cite{Platt:1999}). Thus,
let $q_u$ and $q_v$ be the probabilistic sentiment predictions for $u$
and $v$ on a given bill.  The probability that $u$ and $v$ agree on
the bill is $q_u q_v + (1-q_u) (1-q_v)$, and we define the probability
$p_e$ of a positive sign on the edge $e=\{u,v\}$ as the average
agreement probability over all bills that $u$ and $v$ co\hyp voted on.

\revision{
For instance, the speech containing the sentence,
\textit{`Mr.\ Speaker, I do rise today in strong support of H.R.\ 810,'}
receives a probability of 98\% of expressing a positive opinion on H.R.\ (i.e., House of Representatives) bill 810,
whereas the prediction for the speech containing the words, 
\textit{`Therefore, I urge my colleagues to vote against both H.R.\ 810 and H.R.\ 2520,'}
is only 1\%.
Hence, the edge between the two respective speakers has a probability of
$98\% \times 1\% + 2\% \times 99\% = 3\%$
of being positive.
}

%
%
%
%
%

\subsection{Results}
\label{sec:Convote results}

\begin{figure}
 \centering
 	\subfigure[AUC/ROC]{
	    \hspace{-2mm}
	    \includegraphics[scale=1]{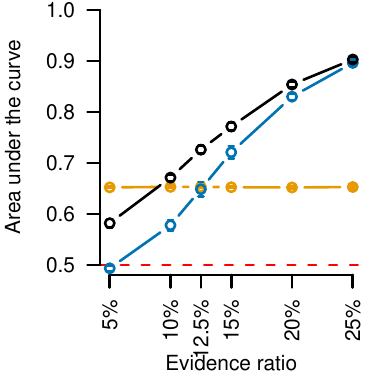}
		\label{fig:auc-roc_convote}
	}
 	\subfigure[AUC/negPR]{
	    \includegraphics[scale=1]{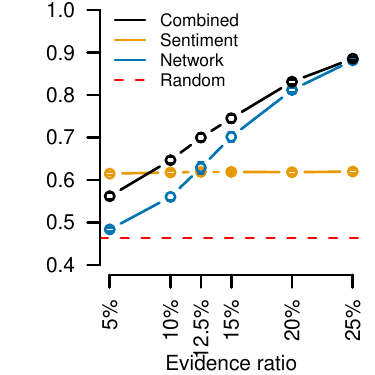}
		\label{fig:auc-pr_neg_convote}
	}
     \vspace{-5mm}
\caption{
	AUC as function of evidence ratio (Convote), with standard errors.
	}
 \label{fig:auc_convote}
\end{figure}

\Figref{fig:auc_convote} summarizes our results. As in the Wikipedia
experiments, we report AUCs as a function of the 
evidence ratio.  The sentiment model alone (yellow)
achieves an AUC/ROC (AUC/negPR) of 0.65 (0.62), well above the random
baselines at 0.5 (0.46).  The network\hyp only model (blue) performs
much worse at the start, but it surpasses the sentiment
model even with just 12.5\% of the edges as evidence, a
reflection of the dense, high\hyp quality network structure with many
triangles.  When we
combine the sentiment and network models (black), we consistently
see the best results, with the largest gains in the
realistic scenario where there is little evidence.

%
%
%
%

Eventually, the network\hyp only model catches up to the combined model,
simply because it reaches an upper bound on performance given
available evidence. This owes mainly to the fact that, because we
derived the person--person signs from person--item signs, only
triangles with an even number of negative edges arise with noticeable
frequency.  To see why, suppose the corpus contained speeches about just
one bill. In a triangle consisting of nodes $u$, $v$, and $w$, if $u$
agreed with $v$ and with $w$, then $v$ and $w$ must agree as well.
(The fact that we have multiple bills in the corpus opens up the
possibility for additional triangle types, but they rarely arise in the
data.)
This constrains the solution space and makes the problem easier
than in the case of Wikipedia, where all triangle types are possible.

\begin{figure}
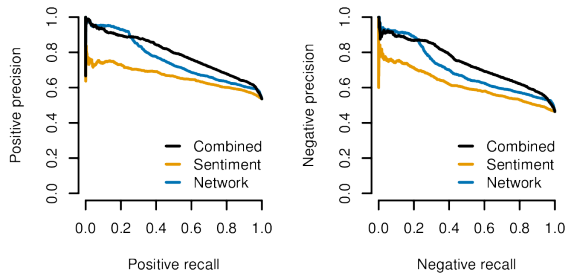

 \centering
 	\subfigure[Positive signs]{
	    \hspace{-2mm}
	    \includegraphics[width=.45\linewidth]{{{FIG/convote/prec-rec_pos_obsRatio=0.15_discard=0}}}
		\label{fig:pr_pos_convote}
	}
 	\subfigure[Negative signs]{
	    \includegraphics[width=.45\linewidth]{{{FIG/convote/prec-rec_neg_obsRatio=0.15_discard=0}}}
		\label{fig:pr_neg_convote}
	}
    \vspace{-3mm}
 \caption{
    \hspace{-2mm}
 	Precision\slash recall (Convote, evidence ratio 15\%).
	}
 \label{fig:pr_convote}
    \vspace{-2mm}
\end{figure}

Our plots so far have summarized precision--recall curves by measuring
AUC.  Here, it is also informative to inspect a concrete PR curve, as
in \figref{fig:pr_convote}, which shows all the values at 15\%
evidence ratio. The network\hyp only model (blue) achieves very high
precision up to a recall of about 0.20, where there is a sudden drop.
The reason is that, according to the above argument about possible triangle
types, the model can be very certain about some edges (e.g., because it is
the only non\hyp evidence edge in a triangle, making only a single triangle
type possible), which causes the plateau for low recall.
The combined model matches the precision on the
plateau, but also maintains significantly higher precision as the
network\hyp only model starts to do more poorly: even if an edge $e$ is not
fully determined by surrounding evidence, the sentiment model might still
give strong signals for $e$ itself and its neighbors, such that the
above reasoning effectively still applies.

\section{Discussion}
\label{sec:Discussion}

We developed a model that synthesizes textual and social\hyp network
information to jointly predict the polarity of person-to-person
evaluations, and we assessed this model in two datasets. Both involve
communal decision making, where people's attitudes and opinions of each
other have profound social consequences, but they are very
different. In the Wikipedia corpus, the sentiment signal is strong
because of established community norms for how to convey one's
opinions, but the network is sparse.  In the Convote corpus, the
network is determined by fully observed voting patterns, making it
strong, but the speech texts themselves only indirectly and noisily convey
person-to-person opinions. In both cases, our method excels because it
is adaptive: it learns from the data how best to combine the two
signals.

Our model's adaptivity is important for real-world applications, where
one is unlikely to know ahead of time which signals are most
trustworthy.
We envision the following use-case. One extracts a coherent subgraph of
the network of interest, perhaps using one of our sampling methods
(\figref{fig:train_test})
and annotates its edges for their evaluativity. Then, in conjunction
with a sentiment model (out-of-the-box or specially trained), one
trains our combined model and uses it to predict new edge labels in the
network. In this setting, the sentiment model might be unreliable, and
one might have the time and resources to label only a small fraction
of the edges. Individually, the network and sentiment models would
likely perform poorly; in bringing the two together, our single model
of joint inference could still excel.

{\small
\xhdr{Acknowledgements}
This research has been supported in part by NSF
IIS-1016909,              
CNS-1010921,            
IIS-1149837,       
IIS-1159679;              
ARO MURI;               
DARPA SMISC,         
GRAPHS;          
ONR N00014-13-1-0287;   
PayPal;                     
Docomo;                    
and
Volkswagen.              
Robert West has been supported by a Facebook and a Stanford Graduate Fellowship.
}

\appendix
\section{Proof sketch of Theorem~\ref{thm:np-completeness}}
\label{sec:proof}

Due to space constraints, we only give a proof sketch here; the full proof is
available online \cite{project-website}.

\begin{figure}
 \centering
    \includegraphics[scale=.66]{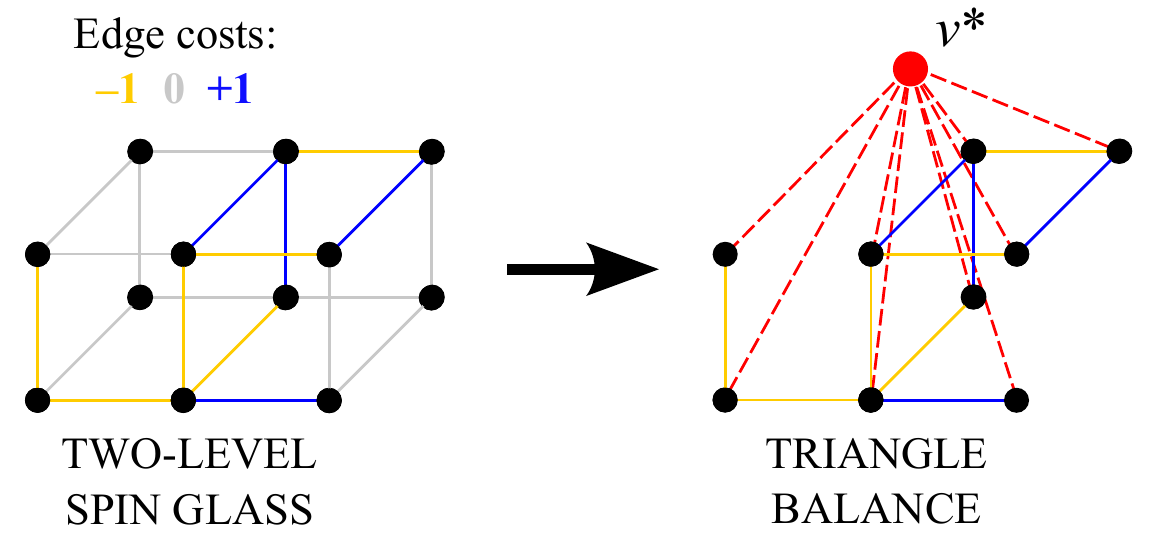}
    \vspace{-3mm}
 \caption{
    Reduction from \TLSG{} to \PROBLEM.
    }
 \label{fig:reduction}
\end{figure}

\begin{proof}[Proof sketch]
By reduction from \TLSG{} (\TLSGshort), a problem known to be NP\hyp hard \cite{barahona1982computational}.
An instance of \TLSGshort{} consists of vertices $V$ arranged in two 2D grids, one stacked above the other, with edges $E$ between nearest neighbors, and with an edge cost $c_{uv} \in \{-1,0,+1\}$ associated with each edge $\{u,v\}$ (see Fig.~\ref{fig:reduction} for a small instance).
Given such an instance, \TLSGshort{} asks for vertex signs $x \in \{-1,+1\}^{|V|}$ that minimize the total energy $H(x) = -\sum_{\{u,v\} \in E} c_{uv}\, x_u\, x_v$.
The crucial observation is that \TLSGshort{} defines vertex costs (implicitly all-zero) and edge costs, and asks for vertex signs, whereas \PROBLEM{} defines edge costs and triangle costs, and asks for edge signs.
That is, vertices (edges) in \TLSGshort{} correspond to edges (triangles) in \PROBLEM{}, and our proposed reduction transforms an original \TLSGshort{} instance into a \PROBLEM{} instance in which each edge corresponds to exactly one original vertex, and each triangle to exactly one original edge.
As shown in Fig.~\ref{fig:reduction}, which depicts the reduction schematically, this is achieved by introducing a new vertex $v^*$ that is connected to each original vertex and thus creates a triangle for each original edge.
The full proof \cite{project-website} shows how the edge and triangle costs can be constructed such that each optimal solution to the \TLSGshort{} instance corresponds to an optimal solution to the \PROBLEM{} instance, and \textit{vice versa}.
\end{proof}

\bibliographystyle{acl2012}
\bibliography{bibliography}

\begin{thebibliography}{}

\bibitem[\protect\citename{Agrawal \bgroup et al.\egroup
  }2003]{Agrawal-etal:2003}
Rakesh Agrawal, Sridhar Rajagopalan, Ramakrishnan Srikant, and Yirong Xu.
\newblock 2003.
\newblock Mining newsgroups using networks arising from social behavior.
\newblock In {\em Proceedings of the 12th International Conference on World
  Wide Web}.

\bibitem[\protect\citename{Anand \bgroup et al.\egroup }2011]{Anand-etal:2011}
Pranav Anand, Marilyn Walker, Rob Abbott, Jean~E. Fox~Tree, Robeson Bowmani,
  and Michael Minor.
\newblock 2011.
\newblock Cats rule and dogs drool! {C}lassifying stance in online debate.
\newblock In {\em Proceedings of the 2nd Workshop on Computational Approaches
  to Subjectivity and Sentiment Analysis}.

\bibitem[\protect\citename{Bach \bgroup et al.\egroup }2013]{bach2013hinge}
Stephen Bach, Bert Huang, Ben London, and Lise Getoor.
\newblock 2013.
\newblock Hinge-loss {M}arkov random fields: Convex inference for structured
  prediction.
\newblock In {\em Proceedings of the 29th Conference on Uncertainty in
  Artificial Intelligence}.

\bibitem[\protect\citename{Barahona}1982]{barahona1982computational}
Francisco Barahona.
\newblock 1982.
\newblock On the computational complexity of {I}sing spin glass models.
\newblock {\em Journal of Physics A: Mathematical and General},
  15(10):3241--3253.

\bibitem[\protect\citename{Blitzer \bgroup et al.\egroup
  }2007]{Blitzer:Dredze:Pereira:2007}
John Blitzer, Mark Dredze, and Fernando Pereira.
\newblock 2007.
\newblock Biographies, bollywood, boom-boxes and blenders: Domain adaptation
  for sentiment classification.
\newblock In {\em Proceedings of the 45th Annual Meeting of the Association of
  Computational Linguistics}.

\bibitem[\protect\citename{Boyd \bgroup et al.\egroup
  }2011]{boyd2011distributed}
Stephen Boyd, Neal Parikh, Eric Chu, Borja Peleato, and Jonathan Eckstein.
\newblock 2011.
\newblock Distributed optimization and statistical learning via the alternating
  direction method of multipliers.
\newblock {\em Foundations and Trends in Machine Learning}, 3(1):1--122.

\bibitem[\protect\citename{Broecheler \bgroup et al.\egroup
  }2010]{broecheler:uai10}
Matthias Broecheler, Lilyana Mihalkova, and Lise Getoor.
\newblock 2010.
\newblock Probabilistic similarity logic.
\newblock In {\em Proceedings of the 26th Conference on Uncertainty in
  Artificial Intelligence}.

\bibitem[\protect\citename{Cartwright and Harary}1956]{Cartwright:Harary:1956}
Dorwin Cartwright and Frank Harary.
\newblock 1956.
\newblock Structure balance: A generalization of {H}eider's theory.
\newblock {\em Psychological Review}, 63(5):277--293.

\bibitem[\protect\citename{Dunbar}2004]{Dunbar:2004}
Robin~I. Dunbar.
\newblock 2004.
\newblock Gossip in evolutionary perspective.
\newblock {\em Review of General Psychology}, 8(2):100--110.

\bibitem[\protect\citename{Feldman~Barrett and
  Russell}1998]{Feldman:Russell:1998}
Lisa Feldman~Barrett and James~A. Russell.
\newblock 1998.
\newblock Independence and bipolarity in the structure of affect.
\newblock {\em Journal of Personality and Social Psychology}, 74(4):967--984.

\bibitem[\protect\citename{Heider}1946]{Heider:1946}
Fritz Heider.
\newblock 1946.
\newblock Attitudes and cognitive organization.
\newblock {\em The Journal of Psychology}, 21(1):107--112.

\bibitem[\protect\citename{Hu \bgroup et al.\egroup }2013]{hu2013exploiting}
Xia Hu, Lei Tang, Jiliang Tang, and Huan Liu.
\newblock 2013.
\newblock Exploiting social relations for sentiment analysis in microblogging.
\newblock In {\em Proceedings of the 6th ACM International Conference on Web
  Search and Data Mining}.

\bibitem[\protect\citename{Huang \bgroup et al.\egroup
  }2013]{huang2013flexible}
Bert Huang, Angelika Kimmig, Lise Getoor, and Jennifer Golbeck.
\newblock 2013.
\newblock A flexible framework for probabilistic models of social trust.
\newblock In {\em Proceedings of the 2013 International Social Computing,
  Behavioral-Cultural Modeling and Prediction Conference}.

\bibitem[\protect\citename{Jurafsky \bgroup et al.\egroup
  }2014]{Jurafsky-etal:2014}
Dan Jurafsky, Victor Chahuneau, Bryan~R. Routledge, and Noah~A. Smith.
\newblock 2014.
\newblock Narrative framing of consumer sentiment in online restaurant reviews.
\newblock {\em First Monday}, 19(4--7).

\bibitem[\protect\citename{Koller and Friedman}2009]{koller2009probabilistic}
Daphne Koller and Nir Friedman.
\newblock 2009.
\newblock {\em Probabilistic Graphical Models: Principles and Techniques}.
\newblock MIT Press.

\bibitem[\protect\citename{Kunegis \bgroup et al.\egroup
  }2013]{kunegis2013added}
J{\'e}r{\^o}me Kunegis, Julia Preusse, and Felix Schwagereit.
\newblock 2013.
\newblock What is the added value of negative links in online social networks?
\newblock In {\em Proceedings of the 22nd International Conference on World
  Wide Web}.

\bibitem[\protect\citename{Leskovec \bgroup et al.\egroup
  }2010a]{leskovec2010predicting}
Jure Leskovec, Daniel Huttenlocher, and Jon Kleinberg.
\newblock 2010a.
\newblock Predicting positive and negative links in online social networks.
\newblock In {\em Proceedings of the 19th International Conference on World
  Wide Web}.

\bibitem[\protect\citename{Leskovec \bgroup et al.\egroup
  }2010b]{leskovec2010signed}
Jure Leskovec, Daniel Huttenlocher, and Jon Kleinberg.
\newblock 2010b.
\newblock Signed networks in social media.
\newblock In {\em Proceedings of the SIGCHI Conference on Human Factors in
  Computing Systems}.

\bibitem[\protect\citename{Lin \bgroup et al.\egroup }2006]{Lin-etal:2006}
Wei-Hao Lin, Theresa Wilson, Janyce Wiebe, and Alexander Hauptmann.
\newblock 2006.
\newblock Which side are you on? {I}dentifying perspectives at the document and
  sentence levels.
\newblock In {\em Proceedings of the 10th Conference on Computational Natural
  Language Learning}.

\bibitem[\protect\citename{Ma \bgroup et al.\egroup }2011]{ma2011recommender}
Hao Ma, Dengyong Zhou, Chao Liu, Michael~R Lyu, and Irwin King.
\newblock 2011.
\newblock Recommender systems with social regularization.
\newblock In {\em Proceedings of the 4th ACM International Conference on Web
  Search and Data Mining}.

\bibitem[\protect\citename{Malouf and Mullen}2008]{Malouf:Mullen:2008}
Robert Malouf and Tony Mullen.
\newblock 2008.
\newblock Taking sides: User classification for informal online political
  discourse.
\newblock {\em Internet Research}, 18(2):177--190.

\bibitem[\protect\citename{Pang and Lee}2004]{PangLee04}
Bo~Pang and Lillian Lee.
\newblock 2004.
\newblock A sentimental education: Sentiment analysis using subjectivity
  summarization based on minimum cuts.
\newblock In {\em Proceedings of the 42nd Annual Meeting of the Association for
  Computational Linguistics}.

\bibitem[\protect\citename{Pang and Lee}2008]{PangLee08}
Bo~Pang and Lillian Lee.
\newblock 2008.
\newblock Opinion mining and sentiment analysis.
\newblock {\em Foundations and Trends in Information Retrieval}, 2(1):1--135.

\bibitem[\protect\citename{Platt}1999]{Platt:1999}
John Platt.
\newblock 1999.
\newblock Probabilistic outputs for support vector machines and comparisons to
  regularized likelihood methods.
\newblock {\em Advances in Large Margin Classifiers}, 10(3):61--74.

\bibitem[\protect\citename{Recasens \bgroup et al.\egroup
  }2013]{Recasens-etal:2013}
Marta Recasens, Cristian Danescu-Niculescu-Mizil, and Dan Jurafsky.
\newblock 2013.
\newblock Linguistic models for analyzing and detecting biased language.
\newblock In {\em Proceedings of the 51st Annual Meeting of the Association for
  Computational Linguistics}.

\bibitem[\protect\citename{Rubin and Talerico}2009]{Rubin:Talerico:2009}
David~C. Rubin and Jennifer~M. Talerico.
\newblock 2009.
\newblock A comparison of dimensional models of emotion.
\newblock {\em Memory}, 17(8):802--808.

\bibitem[\protect\citename{Russell}1980]{Russell80}
James~A. Russell.
\newblock 1980.
\newblock A circumplex model of affect.
\newblock {\em Journal of Personality and Social Psychology}, 39(6):1161--1178.

\bibitem[\protect\citename{Somasundaran and
  Wiebe}2010]{Somasundaran:Wiebe:2010}
Swapna Somasundaran and Janyce Wiebe.
\newblock 2010.
\newblock Recognizing stances in ideological on-line debates.
\newblock In {\em Proceedings of the {NAACL} {HLT} 2010 Workshop on
  Computational Approaches to Analysis and Generation of Emotion in Text}.

\bibitem[\protect\citename{Tan \bgroup et al.\egroup }2011]{Tan-etal:2011}
Chenhao Tan, Lillian Lee, Jie Tang, Long Jiang, Ming Zhou, and Ping Li.
\newblock 2011.
\newblock User-level sentiment analysis incorporating social networks.
\newblock In {\em Proceedings of the 17th {ACM} {SIGKDD} International
  Conference on Knowledge Discovery and Data Mining}.

\bibitem[\protect\citename{Tang \bgroup et al.\egroup
  }2013]{Tang:2013:EHE:2433396.2433405}
Jiliang Tang, Huiji Gao, Xia Hu, and Huan Liu.
\newblock 2013.
\newblock Exploiting homophily effect for trust prediction.
\newblock In {\em Proceedings of the 6th ACM International Conference on Web
  Search and Data Mining}.

\bibitem[\protect\citename{Thomas \bgroup et al.\egroup
  }2006]{Thomas:Pang:Lee:2006}
Matt Thomas, Bo~Pang, and Lillian Lee.
\newblock 2006.
\newblock Get out the vote: Determining support or opposition from
  {C}ongressional floor-debate transcripts.
\newblock In {\em Proceedings of the 2006 Conference on Empirical Methods in
  Natural Language Processing}.

\bibitem[\protect\citename{West}2014]{project-website}
Robert West.
\newblock 2014.
\newblock Supplementary material.
\newblock Online.
\newblock http://infolab.stanford.edu/$\sim$west1/TACL2014/.

\bibitem[\protect\citename{Wilson \bgroup et al.\egroup }2005]{Wilson-etal05}
Theresa Wilson, Janyce Wiebe, and Paul Hoffmann.
\newblock 2005.
\newblock Recognizing contextual polarity in phrase-level sentiment analysis.
\newblock In {\em Proceedings of Human Language Technology Conference and
  Conference on Empirical Methods in Natural Language Processing}.

\bibitem[\protect\citename{Yano \bgroup et al.\egroup }2010]{Yano-etal10}
Tae Yano, Philip Resnik, and Noah~A. Smith.
\newblock 2010.
\newblock Shedding (a thousand points of) light on biased language.
\newblock In {\em Proceedings of the {NAACL}-{HLT} Workshop on Creating Speech
  and Language Data With {A}mazon's {M}echanical {T}urk}.

\bibitem[\protect\citename{Yu \bgroup et al.\egroup
  }2008]{Yu:Kaufmann:Diermeier:2008}
Bei Yu, Stefan Kaufmann, and Daniel Diermeier.
\newblock 2008.
\newblock Classifying party affiliation from political speech.
\newblock {\em Journal of Information Technology and Politics}, 5(1):33--48.

\end{thebibliography}

\end{document}